\documentclass[preprint,showpacs,preprintnumbers,amsmath,amssymb]{revtex4}

\usepackage{graphicx}

\begin{document}

\title{
Collinear solution 
to the general relativistic three-body problem 
}
\author{Kei Yamada}
\author{Hideki Asada} 
\affiliation{
Faculty of Science and Technology, Hirosaki University,
Hirosaki 036-8561, Japan} 

\date{\today}

\begin{abstract}
The three-body problem is reexamined 
in the framework of general relativity. 
The Newtonian three-body problem admits 
{\it Euler's collinear solution}, 
where three bodies move around the common center of 
mass with the same orbital period and always line up. 
The solution is unstable. 
Hence it is unlikely that such a simple configuration would 
exist owing to general relativistic forces 
dependent not only on the masses but also on 
the velocity of each body. 
However, we show that the collinear solution remains true 
with a correction to the spatial separation between masses. 
Relativistic corrections to the Sun-Jupiter Lagrange points 
$L_1$, $L_2$ and $L_3$ are also evaluated. 
\end{abstract}


\maketitle

\noindent \emph{Introduction.--- } 
The three-body problem in the Newton gravity belongs 
among classical problems in astronomy and physics 
(e.g, \cite{Danby,Goldstein}). 
In 1765, Euler found a collinear solution for 
the restricted three-body problem, 
where one of three bodies is a test mass. 
Soon later, his solution was extended for 
a general three-body problem by Lagrange, 
who also found an equilateral triangle solution 
in 1772. 
Now, the solutions for the restricted three-body problem 
are called Lagrange points $L_1, L_2, L_3, L_4$ and $L_5$, 
which are described 
in textbooks of classical mechanics \cite{Goldstein}. 
SOHO and WMAP launched by NASA are in operation 
at the Sun-Earth $L_1$ and $L_2$, respectively. 
LISA pathfinder is planned to go to $L_1$. 
Lagrange points have recently attracted renewed interests 
for relativistic astrophysics \cite{SM,Schnittman}, 
where they have discussed the gravitational radiation 
reaction on $L_4$ and $L_5$ by numerical methods. 
As a pioneering work, Nordtvedt pointed out that 
the location of the triangular points is very sensitive 
to the ratio of the gravitational mass to the inertial one 
\cite{Nordtvedt}. 
Along this course, it is interesting as a gravity experiment 
to discuss the three-body coupling terms at the post-Newtonian 
order, 
because some of the terms are proportional to a product of 
three masses as $M_1 \times M_2 \times M_3$. 
Such a term appears only for relativistic three (or more) 
body systems: 
For a relativistic binary with two masses $M_1$ and $M_2$, 
$M_1^2 M_2$ and $M_1 M_2^2$ exist but 
such three mass products do not. 
For a Newtonian three-body system, we have 
only the terms proportional to 
$M_1 M_2$, $M_2 M_3$ and $M_3 M_1$. 
The relativistic periastron advance of the Mercury 
is detected only after much larger shifts due to 
Newtonian perturbations by other planets such as 
the Venus and Jupiter are taken into account 
in the astrometric data analysis. 
In this sense, effects by the three body coupling 
are worthy to investigate.

After efforts to find a general solution, 
Poincare proved that 
it is impossible to describe all the solutions 
to the three-body problem even for the $1/r$ potential. 
Namely, we cannot analytically obtain all the solutions. 
Nevertheless, the number of new solutions is increasing \cite{Marchal}.   
Therefore, the three-body problem remains unsettled 
even for Newton gravity. 

The theory of general relativity is currently the most successful 
gravitational theory describing the nature of space and time, 
and well confirmed by observations. 
Especially, it has passed ``classical'' tests, 
such as the deflection of light, the perihelion shift 
of Mercury and the Shapiro time delay, and also a systematic test 
using the remarkable binary pulsar ``PSR 1913+16'' \cite{Will}. 
It is worthwhile to examine the three-body (or more generally, N-body) 
problem in general relativity. 
However, it is difficult to work out in general relativity 
compared with Newton gravity, 
because the Einstein equation is much more complicated \cite{MTW} 
(even for a two-body system \cite{PW,FJS,Blanchet,AF}).  
So far, most of post-Newtonian works have focused on 
either compact binaries 
for an application to gravitational waves astronomy 
or 
N-body equation of motion (and coordinate systems) 
in the weak field such as the solar system (e.g. \cite{Brumberg}). 
In addition, future space astrometric missions 
such as SIM and GAIA 
\cite{SIM,GAIA,JASMINE}
require a general relativistic modeling of 
the solar system within the accuracy of a micro arc-second 
\cite{Klioner}. 
Furthermore, a binary plus a third body have been discussed 
also for perturbations of gravitational waves induced by the third body 
\cite{ICTN,Wardell,CDHL,GMH}. 

The Newtonian three-body problem admits 
{\it Euler's collinear solution}, 
where three bodies move around the common center of 
mass with the same orbital period and always line up. 
The solution is unstable 
against small displacements. 
Hence it is unlikely that such a simple configuration would 
exist owing to general relativistic forces 
dependent not only on the masses but also on 
the velocity of each body. 
The line could bend at a certain location of one mass, 
which means a V-shape configuration. 
The above Newtonian instability does not necessarily 
come from small perturbations of acceleration. 
Therefore, it is interesting to ask whether 
the general relativistic gravity in the rather complicated form 
admits a collinear solution or lead to such a V-shape solution. 
We shall also evaluate for the first time 
relativistic corrections 
to $L_1$, $L_2$ and $L_3$ for the Sun-Jupiter system.  

In recent, a {\it choreographic} solution 
has been studied in the framework of general relativity \cite{ICA}. 
Here, a solution is called choreographic in the celestial mechanics, 
if every massive particles move periodically in a single closed
orbit. 
As a choreographic solution, the figure-eight one was found 
first by Moore and rediscovered 
with its existence proof by Chenciner and Montgomery 
\cite{Moore,CM,GMF,Simo,Montgomery05}. 
The solution was shown to remain true 
at the first post-Newtonian \cite{ICA} 
and also the second post-Newtonian orders \cite{LN}. 
Such an unexpected feature may be found 
in the collinear solution. 

This paper is organized as follows. 
First, we briefly summarize a usual treatment 
of Euler's collinear solution. 
Next, we extend the formulation to the post-Newtonian case 
by treating the Einstein-Infeld-Hoffman 
equation of motion.  
We take the units of $G=c=1$.

\noindent \emph{Newtonian Euler's collinear solution.---} 
Let us briefly summarize the derivation of 
the Euler's collinear solution for the circular three-body problem 
in Newton gravity. 
We consider Euler's solution, for which  
each mass moves around their common center 
of mass denoted as $\mbox{\boldmath $X$}_G$ 
with a constant angular velocity $\omega$. 
Hence, it is convenient to use the corotating frame 
with the same angular velocity $\omega$. 
We choose an orbital plane normal to the total angular momentum 
as the $x-y$ plane in such a corotating frame. 
We locate all the three bodies along a single line, 
along which we take the $x$-coordinate. 
The location of each mass $M_I$ $(I=1, 2, 3)$ is written as
$\mbox{\boldmath $X$}_I \equiv (x_I, 0)$. 
Without loss of generality, 
we assume $x_3 < x_2 < x_1$. 
Let $R_I$ define the relative position 
of each mass with respective to the center of mass 
$\mbox{\boldmath $X$}_G \equiv (x_G, 0)$, 
namely $R_I \equiv x_I - x_G$ 
($R_I \neq |\mbox{\boldmath $X$}_I|$ unless $x_G = 0$). 
We choose $x=0$ between $M_1$ and $M_3$. 
We thus have $R_3 < R_2 < R_1$, $R_3 < 0$ and $R_1 > 0$. 

\begin{figure}[t]
\includegraphics[width=15cm]{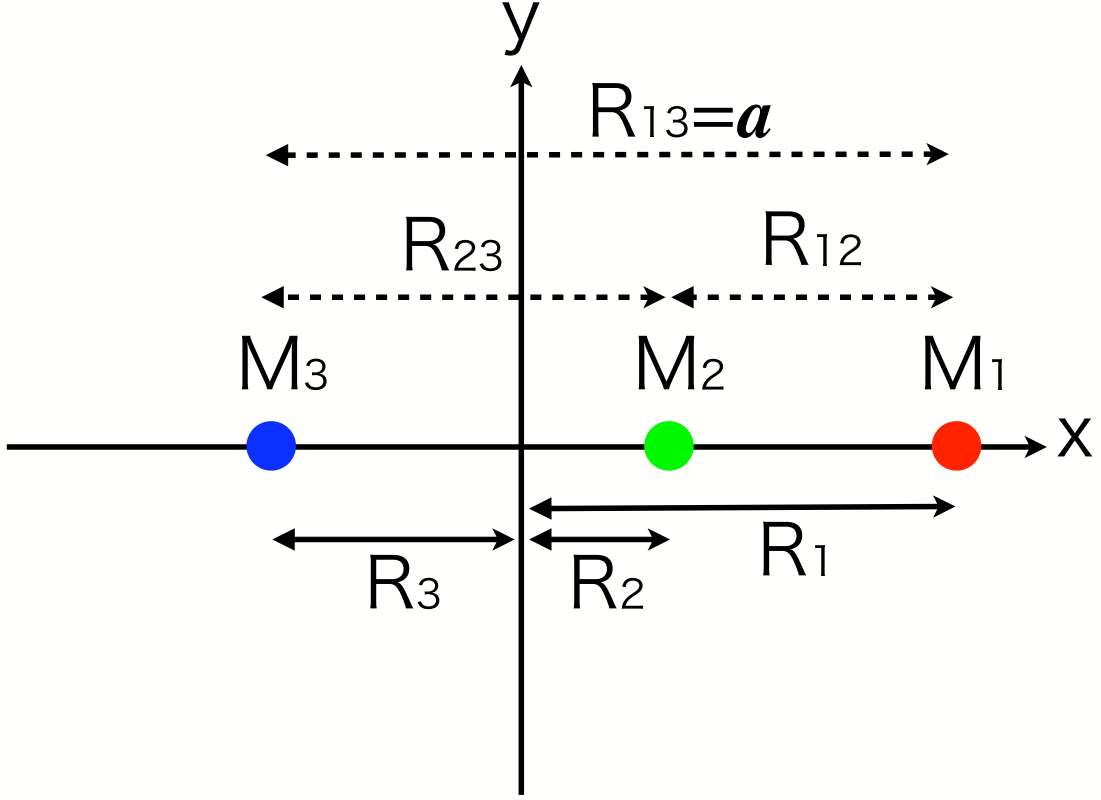}
\caption{ 
(color online). 
Schematic figure for a classical configuration of three masses 
denoted by $M_1$ (red), $M_2$ (green) and $M_3$ (blue), 
which represent a Newtonian collinear solution. 
The filled disks denote each mass at $t=0$. 
Definitions of $a$, $R_I$ and $R_{IJ}$ are also indicated. 
}
\label{f1}
\end{figure}

It is convenient to define a ratio as 
$R_{23} / R_{12}= z$, which is an important variable 
in the following formulation. 
Then we have $R_{13} = (1+z) R_{12}$. 
The equation of motion becomes 
\begin{eqnarray}
R_1 \omega^2 &=& \frac{M_2}{R_{12}^2} + \frac{M_3}{R_{13}^2} , 
\label{EOM-M1-N}
\\
R_2 \omega^2 &=& -\frac{M_1}{R_{12}^2} + \frac{M_3}{R_{23}^2} , 
\label{EOM-M2-N}
\\
R_3 \omega^2 &=& -\frac{M_1}{R_{13}^2} - \frac{M_2}{R_{23}^2} , 
\label{EOM-M3-N} 
\end{eqnarray}
where we define 
\begin{eqnarray}
\mbox{\boldmath $R$}_{IJ} &\equiv& 
\mbox{\boldmath $X$}_{I}-\mbox{\boldmath $X$}_{J} , 
\\
R_{IJ} &\equiv& |\mbox{\boldmath $R$}_{IJ}| . 
\end{eqnarray}
Figure $\ref{f1}$ shows a classical configuration at $t=0$.  
At $t=T_N/2$, this configuration is rotated by $\pi$ radian, 
where $T_N$ denotes the Newtonian orbital period.

First, we subtract Eq. ($\ref{EOM-M2-N}$) from Eq. ($\ref{EOM-M1-N}$) 
and Eq. ($\ref{EOM-M3-N}$) from Eq. ($\ref{EOM-M2-N}$) 
and use 
$R_{12} \equiv |\mbox{\boldmath $X$}_1 - \mbox{\boldmath $X$}_2|$ 
and 
$R_{23} \equiv |\mbox{\boldmath $X$}_2 - \mbox{\boldmath $X$}_3|$. 
Such a subtraction procedure will be useful 
also at the post-Newtonian order, 
because we can avoid directly using  
the post-Newtonian center of mass \cite{MTW,LL}. 
Next, we compute a ratio between them to delete $\omega^2$. 
Hence we obtain a fifth-order equation as 
\begin{equation}
(M_1+M_2) z^5 + (3M_1+2M_2)z^4 + (3M_1+M_2)z^3 
- (M_2+3M_3)z^2 - (2M_2+3M_3)z - (M_2+M_3) = 0. 
\label{5th}
\end{equation}
Now we have a condition as $z>0$. 
Descartes' rule of signs (e.g., \cite{Waerden}) 
states that the number of positive roots either equals to 
that of sign changes in coefficients of a polynomial or 
less than it by a multiple of two. 
According to this rule, 
Eq. ($\ref{5th}$) has the only positive root $z>0$, 
though such a fifth-order equation cannot be solved 
in algebraic manners as shown by Galois (e.g., \cite{Waerden}). 
After obtaining $z$, one can substitute it into a difference, 
for instance between Eqs. ($\ref{EOM-M1-N}$) and ($\ref{EOM-M3-N}$). 
Hence we get $\omega$.

\noindent \emph{post-Newtonian collinear solution.---} 
In the previous 
part, 
the motion of massive bodies 
follows the Newtonian equation of motion. 
In order to include the dominant part of general relativistic effects, 
we take account of the terms at the first post-Newtonian order. 
Namely, the massive bodies obey  
the Einstein-Infeld-Hoffman (EIH) equation of motion as \cite{MTW,LL} 
\begin{eqnarray}
\frac{d \mbox{\boldmath $v$}_K}{dt} 
&=& \sum_{A \neq K} \mbox{\boldmath $R$}_{AK} 
\frac{M_A}{R_{AK}^3} 
\left[
1 - 4 \sum_{B \neq K} \frac{M_B}{R_{BK}} 
- \sum_{C \neq A} \frac{M_C}{R_{CA}} 
\left( 1 - 
\frac{\mbox{\boldmath $R$}_{AK} \cdot \mbox{\boldmath $R$}_{CA}}
{2R_{CA}^2} \right) 
\right.
\nonumber\\
&&
\left. 
~~~~~~~~~~~~~~~~~~~~~
+ v_K^2 + 2v_A^2 - 4\mbox{\boldmath $v$}_A \cdot \mbox{\boldmath $v$}_K 
- \frac32 \left( 
\mbox{\boldmath $v$}_A \cdot \mbox{\boldmath $n$}_{AK} \right)^2 
\right]
\nonumber\\
&&
- \sum_{A \neq K} (\mbox{\boldmath $v$}_A - \mbox{\boldmath $v$}_K) 
\frac{M_A \mbox{\boldmath $n$}_{AK} \cdot 
(3 \mbox{\boldmath $v$}_A - 4 \mbox{\boldmath $v$}_K)}{R_{AK}^2} 
\nonumber\\
&&
+ \frac72 \sum_{A \neq K} \sum_{C \neq A} 
\mbox{\boldmath $R$}_{CA} 
\frac{M_A M_C}{R_{AK} R_{CA}^3} , 
\label{EIH-EOM}
\end{eqnarray}
where $\mbox{\boldmath $v$}_I$ denotes the velocity of each mass 
in an inertial frame 
and we define 
\begin{eqnarray}
\mbox{\boldmath $n$}_{IJ}&\equiv&
\frac{\mbox{\boldmath $R$}_{IJ}}{R_{IJ}} .
\end{eqnarray}
Here, the middle term
with vector $(\mbox{\boldmath $v$}_A - \mbox{\boldmath $v$}_K)$ 
has a zero coefficient for the circular collinear case, while 
the remaining accelerations are radial. 

We obtain a lengthy form of the equation of motion for each body. 
By subtracting the post-Newtonian equation of motion 
for $M_3$ from that for $M_1$ for instance, 
we obtain the equation as 
\begin{equation}
R_{13} \omega^2 = F_N + F_M + F_V \omega^2 , 
\label{EOM-PN}
\end{equation}
where we denote $a \equiv R_{13}$ and  
the Newtonian term $F_N$ 
and the post-Newtonian parts $F_M$ (dependent on the masses only) 
and $F_V$ (velocity-dependent part divided by $\omega^2$) 
are defined as 
\begin{eqnarray}
F_N &=& \frac M{a^2z^2} 
\Biggl[
 (\nu_1 + \nu_3) z^2 + (1 - \nu_1 - \nu_3) (1 + z^2) (1 + z)^2
\Biggr],
\\
F_M &=& - \frac {M^2}{a^3 z^3} 
\Biggl[
 (4 - 4\nu_1 + \nu_3) (1 - \nu_1 - \nu_3)
\notag \\
&& \mspace{65mu}
+ (12 - 7\nu_1 + 3\nu_3) (1 - \nu_1 - \nu_3) z
\notag \\
&& \mspace{65mu}
+ (12 - \nu_1 + \nu_3) (1 - \nu_1 - \nu_3) z^2
\notag \\
&& \mspace{65mu}
+ (8 - 7\nu_1 - 7\nu_3 + 8\nu_1\nu_3 + 3\nu_1^2 + 3\nu_3^2) z^3
\notag\\
&& \mspace{65mu}
+ (12 + \nu_1 - \nu_3) (1 - \nu_1 - \nu_3) z^4
\notag \\
&& \mspace{65mu}
+ (12 + 3\nu_1 - 7\nu_3) (1 - \nu_1 - \nu_3) z^5
\notag \\
&& \mspace{65mu}
+ (4 +  \nu_1 - 4\nu_3) (1 - \nu_1 - \nu_3) z^6 
\Biggr],
\\
F_V &=& \frac M{(1 + z)^2 z^2}
\Biggl[
- \nu_1^2 (1 - \nu_1 - \nu_3)
\notag \\
&& \mspace{91mu}
 - 2\nu_1 (1 + \nu_1 - \nu_3) (1 - \nu_1 - \nu_3) z
\notag \\
&& \mspace{91mu}
+ (2 - 2\nu_1 + \nu_3 + 6\nu_1\nu_3 - 3\nu_3^2 
+ \nu_1^3 - 3\nu_1^2\nu_3 - 3\nu_1\nu_3^2 + \nu_3^3) z^2
\notag \\
&& \mspace{91mu}
+ 2(2 - \nu_1 - \nu_3)
 (1 + \nu_1 + \nu_3 - \nu_1^2 + \nu_1\nu_3 - \nu_3^2) z^3
\notag \\
&& \mspace{91mu}
+ (2 + \nu_1 - 2\nu_3 - 3\nu_1^2 + 6\nu_1\nu_3
+ \nu_1^3 - 3\nu_1^2\nu_3 - 3\nu_1\nu_3^2 + \nu_3^3) z^4
\notag \\
&& \mspace{91mu}
- 2\nu_3 (1 - \nu_1 + \nu_3) (1 - \nu_1 - \nu_3) z^5
\notag \\
&& \mspace{91mu}
- \nu_3^2 (1 - \nu_1 - \nu_3) z^6
\Biggr],
\end{eqnarray} 
respectively. 
Here, we define 
the mass ratio as $\nu_I \equiv M_I/M$ 
for $M \equiv \sum_I M_I$ 
and frequently use $\nu_2=1-\nu_1-\nu_3$. 
It should be noted that 
in this truncated calculation we ignore the second post-Newtonian 
(or higher order) contributions so that 
we can replace, for instance,  
$v_1$ by $R_1 \omega$ (with using the Newtonian $R_1$) 
in post-Newtonian velocity-dependent terms such as $v_1^2$. 

After straightforward but lengthy calculations, 
which are similar to the above Newtonian case, 
we obtain a seventh-order equation as 
\begin{equation}
F(z) \equiv \sum_{k=0}^7 A_k z^k = 0 , 
\label{7th}
\end{equation}
where we define 
\begin{eqnarray}
A_7 &=& \frac Ma 
\Biggl[
- 4 - 2(\nu_1 - 4\nu_3)
+ 2(\nu_1^2 + 2\nu_1\nu_3 - 2\nu_3^2)
- 2\nu_1\nu_3(\nu_1 + \nu_3)
\Biggr],
\label{A7}
\\
A_6 &=& 1 - \nu_3
+ \frac Ma
\Biggl[
- 13 - (10\nu_1 - 17\nu_3)
+ 2(2\nu_1^2 + 8\nu_1\nu_3 - \nu_3^2)
\notag \\
&& \mspace{103mu}
+ 2(\nu_1^3 - 2\nu_1^2\nu_3 - 3\nu_1\nu_3^2 - \nu_3^3)
\Biggr] ,
\label{A6}
\\
A_5 &=& 2 + \nu_1 - 2\nu_3
+ \frac Ma 
\Biggl[
- 15 -(18\nu_1-5\nu_3) + 4(5\nu_1\nu_3 + 4\nu_3^2)
  \notag \\
&& \mspace{148mu}
+ 6(\nu_1^3 - \nu_1\nu_3^2 - \nu_3^3)
\Biggr] ,
\label{A5}
\\
A_4 &=& 1 + 2\nu_1 - \nu_3
+ \frac Ma
\Biggl[
- 6 - 2(5\nu_1 + 2\nu_3)
- 4(2\nu_1^2 - \nu_1\nu_3 - 4\nu_3^2)
\notag \\
&& \mspace{148mu}
+ 2(3\nu_1^3 + \nu_1^2\nu_3 - 2\nu_1\nu_3^2 - 3\nu_3^3)
\Biggr] ,
\label{A4}
\\
A_3 &=& - (1 - \nu_1 + 2\nu_3)
+ \frac Ma 
\Biggl[
6 + 2(2\nu_1 + 5\nu_3)
- 4( 4\nu_1^2 + \nu_1\nu_3 - 2\nu_3^2)
\notag \\
&& \mspace{177mu}
+ 2( 3\nu_1^3 + 2\nu_1^2\nu_3 - \nu_1\nu_3^2 - 3\nu_3^3)
\Biggr] ,
\label{A3}
\\
A_2 &=& - (2 - 2\nu_1 + \nu_3)
+ \frac Ma 
\Biggl[
15 - ( 5\nu_1 - 18\nu_3)
- 4(4\nu_1^2 + 5\nu_1\nu_3)
\notag \\
&& \mspace{177mu}
+ 6( \nu_1^3 + \nu_1^2\nu_3 - \nu_3^3)
\Biggr] ,
\label{A2}
\\
A_1 &=& - (1 - \nu_1)
+ \frac Ma 
\Biggl[
 13 - ( 17\nu_1 - 10\nu_3)
+ 2( \nu_1^2 - 8\nu_1\nu_3 - 2\nu_3^2)
\notag \\
&& \mspace{130mu}
+ 2( \nu_1^3 + 3\nu_1^2\nu_3 + 2\nu_1\nu_3^2 - \nu_3^3)
\Biggr] ,
\label{A1}
\\
A_0 &=& \frac Ma 
\Biggl[
4 - 2( 4\nu_1 - \nu_3)
+ 2(2\nu_1^2 - 2\nu_1\nu_3 - \nu_3^2)
+ 2\nu_1\nu_3(\nu_1 + \nu_3) 
\Biggr] .
\label{A0}
\end{eqnarray}
This seventh-order equation is symmetric for exchanges 
between $\nu_1$ and $\nu_3$, 
only if one makes a change as $z \to 1/z$. 
This symmetry may validate the complicated form of 
each coefficient. 

Once a positive root for Eq. ($\ref{7th}$) is found, 
the root $z$ can be substituted into Eq. ($\ref{EOM-PN}$) 
in order to obtain the angular velocity $\omega$. 

The angular velocity including the post-Newtonian effects 
is obtained from Eq. ($\ref{EOM-PN}$) as 
\begin{equation}
\omega = \omega_N 
\left( 1 + \frac{F_M}{2F_N} + \frac{F_V}{2R_{13}} \right) , 
\label{omega}
\end{equation}
where $\omega_N \equiv (F_N/R_{13})^{1/2}$ 
denotes the angular velocity of the Newtonian collinear orbit. 

\begin{figure}[t]
\includegraphics[width=15cm]{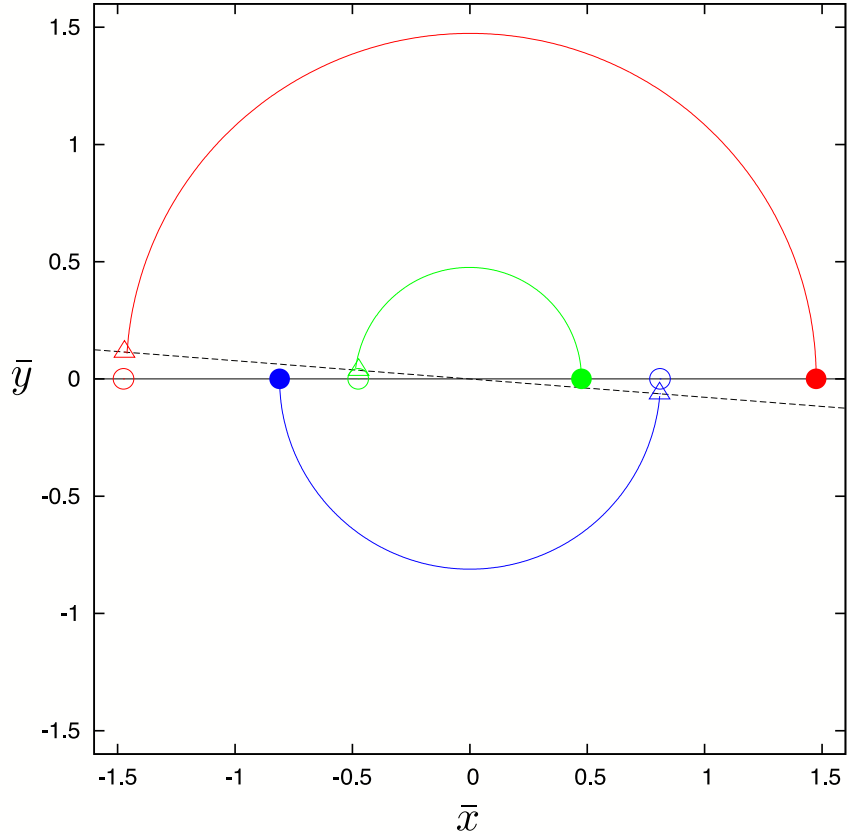}
\caption{
(color online). 
Orbit of each mass representing the post-Newtonian 
collinear solution in an inertial frame. 
We assume 
$M_1\mbox{(red)}:M_2\mbox{(green)}:M_3\mbox{(blue)} = 1:2:3$, 
$R_{12}=1$ and $a/M = 100$.  
The filled disks denote each mass at $t=0$, 
the triangles at $t=T_N/2$, 
and 
the circles at $t=T/2$, 
which can be obtained also by the reflection of 
the filled disks with respect to the $\bar{x}=0$ line. 
}
\label{f2}
\end{figure}

Figure $\ref{f2}$ shows a numerical example 
for $M_1:M_2:M_3 = 1:2:3$, $R_{12}=1$ and $a/M = 100$, 
where the post-Newtonian correction is 
of the order of one percent. 
In this figure, we employ the inertial frame $(\bar{x}, \bar{y})$ 
but not the corotating frame $(x, y)$. 
We assume $x_3<x_2<x_1$ throughout this paper. 
This figure suggests that as an alternative initial condition 
we can assume $x_1<x_2<x_3$, 
which is realized at $t=T/2$ ($T$=orbital period) in this figure. 
It is natural that this is a consequence of the parity symmetry 
in our formulation. 
Numerical calculations for this figure show that 
the relativistic correction in Eq. ($\ref{omega}$) is negative, 
that is $\omega < \omega_N$. 
It should be noted also that the location of each mass at $t=T/2$ 
is advanced compared with that at $t=T_N/2$ 
(a half of the {\it Newtonian} orbital period). 
This may correspond to the periastron advance (in circular orbits). 

We produce this figure in two ways. 
One is that we use our formulation to determine $\omega$ 
and consequently $T$. 
Also $\omega_N$ and $T_N$ are obtained at the Newtonian level. 
Next we rotate the configuration 
by angles $\pi \times (T_N/T)$ and $\pi$, respectively. 
The other is that we directly see the evolution 
of the post-Newtonian system. 
That is, we solve numerically the EIH equation of 
motion until $t=T_N/2$ and $t=T/2$, respectively.
The both methods provide the same plot.  
This agreement may also validate our formulation.

Finally, we focus on the restricted three-body problem 
so that we can put $z = z_N(1+\varepsilon)$ 
for the Newtonian root $z_N$. 
Substitution of this into Eq. ($\ref{7th}$) gives 
the post-Newtonian correction as 
\begin{equation}
\varepsilon = 
- \frac{\sum_k A_{PN k} z_N^k}{\sum_k k A_{N k} z_N^k} , 
\label{varepsilon}
\end{equation}
where $A_{N k}$ and $A_{PN k}$ denote 
the Newtonian and post-Newtonian parts of $A_k$, respectively. 
For a binary system of comparable mass stars, 
the correction $\varepsilon$ is $O(M/a)$.  
This implies that a corrected length is of the order 
of the Schwarzschild radius. 

For the Sun-Jupiter system, general relativistic corrections 
to $L_1$, $L_2$ and $L_3$ become 
$+30$, $-38$, $+1$ [m], respectively, 
where the positive sign is chosen 
along the direction from the Sun to the Jupiter. 
Such corrections suggest a 
potential role
of the general relativistic three (or more) body dynamics 
for high precision astrometry in our solar system 
and perhaps also for gravitational waves astronomy. 
They are very small but may be marginally within the limits 
of the current technology, since 
the Lunar Laser Ranging experiment has successfully measured 
the increasing distance of the Moon $\sim 3.8\mbox{cm/yr}$.

\noindent \emph{Conclusion.--- }
We obtained a general relativistic version of 
Euler's collinear solution for the three-body problem 
at the post-Newtonian order. 
Studying global properties of the seventh-order equation 
that we have derived is left as future work. 

It is interesting also to include higher post-Newtonian 
corrections, especially 2.5PN effects in order to 
elucidate the secular evolution of the orbit 
due to the gravitational radiation reaction at the 2.5PN order. 
One might see probably a shrinking collinear orbit 
as a consequence of a decrease in the total energy and angular momentum, 
if such a radiation reaction effect is included.
This is a testable prediction. 

It may be important also to search other solutions, 
notably a relativistic counterpart of 
the Lagrange's triangle solution 
(so-called $L_4$ and $L_5$ in the restricted three-body problem). 
Clearly it seems much more complicated to obtain 
relativistic corrections to the Lagrange orbit. 

We are grateful to A. Chenciner, Y. Kojima, K. Tanikawa 
for useful conversations. 
This work was supported in part (H.A.) 
by a Japanese Grant-in-Aid 
for Scientific Research from the Ministry of Education, 
No. 21540252.

\end{document}